\documentclass[10pt,letterpaper]{article} 
\usepackage{opex3}

\usepackage{graphicx,ulem}
\usepackage{amsmath,amstext,cite}

\newcommand{\kz}{k}
\newcommand{\kzp}{\kz'}

\newcommand{\bpp}{\kz^{(2)}}
\newcommand{\Dm}{\hat{\mathit{D}}}

\newcommand{\Ein}{{\mathcal E}_{\rm in}}
\newcommand{\neffs}{N_{\rm eff}}%
\newcommand{\neffc}{N_{\rm eff,c}}%
\newcommand{\nshgb}{N_{\rm SHG}}%
\newcommand{\nkerrb}{N_{\rm Kerr}}%
\newcommand{\ld}{L_{\rm D,1}}
\newcommand{\dksr}{\Delta k_{\rm sr}}
\newcommand{\deff}{d_{\rm eff}}

\newcommand{\lgvm}{L_{\rm GVM}}
\newcommand{\lcoh}{L_{\rm coh}}
\newcommand{\kp}{k^{(1)}}
\newcommand{\kpp}{k^{(2)}}
\newcommand{\tin}{T_{\rm in}}
\newcommand{\Iin}{I_{\rm in}}
\newcommand{\imm}{~{\rm mm^{-1}}}
\newcommand{\topt}{\Delta t_{\rm opt}}
\newcommand{\toptc}{\Delta t_{\rm opt}^{\rm corr}}

\newcommand{\Rnl}{{\mathcal R}}
\newcommand{\FT}{{\mathcal F}}
\newcommand{\IFT}{{\mathcal F^{-1}}}
\newcommand{\ie}{{\textit{i.e.}, }}
\newcommand{\ta}{\tau_a}
\newcommand{\tb}{\tau_b}
\newcommand{\tap}{t_a}
\newcommand{\tbp}{t_b}
\newcommand{\tdummy}{s}
\newcommand{\HOD}{HOD }
\begin{document}
\title{Limits to compression with cascaded quadratic soliton compressors} 
\author{M. Bache,$^{1,*}$ O. Bang,$^1$ W. Krolikowski,$^{1,2}$ J. Moses,$^3$ and F. W. Wise$^4$}
\address{$^1$DTU Photonics, 
Technical University of Denmark,
  Bld. 345v, DK-2800 Lyngby, Denmark\\
$^2$Research School of Physical Sciences and Engineering,
Australian National University, Canberra ACT 0200, Australia\\
$^3$Optics and Quantum Electronics Group, Massachusetts Institute of
Technology, Cambridge, MA 02139\\ 
$^4$Department of Applied and Engineering Physics, Cornell
  University, Ithaca, New York 14853}
\email{bache@com.dtu.dk}
\centerline{\small\today}
\begin{abstract}
  We study cascaded quadratic soliton compressors and address the
  physical mechanisms that limit the compression. A nonlocal model is
  derived, and the nonlocal response is shown to have an additional
  oscillatory component in the nonstationary regime when the
  group-velocity mismatch (GVM) is strong. This inhibits efficient
  compression. Raman-like perturbations from the cascaded
  nonlinearity, competing cubic nonlinearities, higher-order
  dispersion, and soliton energy may also limit compression, and
  through realistic numerical simulations we point out when each
  factor becomes important. We find that it is theoretically possible
  to reach the single-cycle regime by compressing high-energy fs
  pulses for wavelengths $\lambda=1.0-1.3~\mu{\rm m}$ in a
  $\beta$-barium-borate crystal, and it requires that the system is in
  the stationary regime, where the phase mismatch is large enough to
  overcome the detrimental GVM effects. However, the simulations show
  that reaching single-cycle duration is ultimately inhibited by
  competing cubic nonlinearities as well as dispersive waves, that
  only show up when taking higher-order dispersion into account.
\end{abstract}


\section{Introduction}
\label{sec:Introduction}

Compression of optical pulses can be achieved by first inducing a
nonlinear phase shift on the pulse by self-phase modulation (SPM) from
the cubic nonlinear response of, e.g., an optical fiber \cite{agrawal:2001}.
The phase shift creates a chirp across the pulse, which means that
compression can subsequently be achieved in a dispersive material
(such as a grating pair). In cubic soliton compressors both the SPM induced
chirp and the compression is achieved in the same
material \cite{mollenauer:1980}. The self-focusing cubic nonlinearity
requires anomalous dispersion to compress the pulse. Due to collapse
problems this limits the energy of the compressed pulse, and the
requirement of anomalous dispersion restricts the accessible
wavelength regime for soliton compression.

Recent progress has shown that cascaded quadratic soliton compressors
(CQSCs) may effectively compress high-energy fs pulses down to
ultra-short few-cycle pulses
\cite{ashihara:2002,ashihara:2004,moses:2005,moses:2006,moses:2007,zeng:2006,xie:2007,bache:2007a,bache:2007}.
Here the nonlinear phase shift is induced due to phase-mismatched
second-harmonic generation (SHG), which acts as a cascaded quadratic
nonlinear process. The pump, or fundamental wave (FW), experiences an
effective SPM from the cyclic energy transfer to the second harmonic
(SH). The advantage is that the effective cubic nonlinearity induced
by the cascading process can be made self-defocusing because it is
controlled by the sign of the phase-mismatch parameter
\cite{desalvo:1992,clausen:1997,liu:1999,ditrapani:2001},
and therefore normal dispersion can be used to compress the FW.

As this compressor scheme exploits an effective self-defocusing cubic
term from cascaded quadratic effects, the compressor will naturally be
affected by the self-focusing cubic nonlinearity inherent to any
transparent material. This detrimental cubic nonlinearity must be
counterbalanced and then exceeded to achieve compression
\cite{liu:1999,ashihara:2002,moses:2006,bache:2007}. Furthermore, the
propagation in a bulk medium having self-focusing cubic nonlinear
response can avoid collapse problems if the self-defocusing cascaded
nonlinearity is strong enough \cite{PhysRevE.55.3555}. Thus, this
compressor works in a bulk configuration even with multi-mJ input
pulse energies \cite{liu:1999}.

\begin{figure}[tb]
  \begin{center}
    \centerline{\includegraphics[width=13cm]{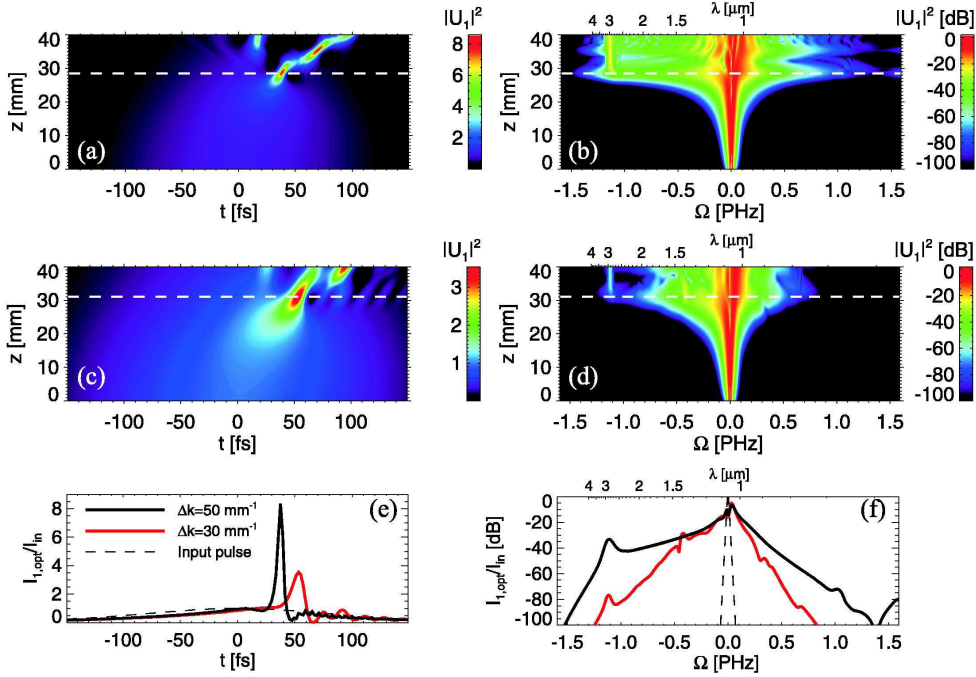}}
\caption{Numerical simulations of soliton compression of a 200 fs FWHM
  pulse in a BBO crystal pumped at $\lambda_1=1064$ nm with a soliton
  number of $\neffs=8$.  (a) and (b) Temporal and spectral components
  of the FW $|U_1|^2=I_1/\Iin$ in the stationary regime ($\Delta
  k=50\imm$, $\Iin=59~{\rm GW/cm^2}$). The pulse is compressed to 6 fs
  at the optimal compression point [$z=z_{\rm opt}$, dashed line, cuts
  in (e) and (f)].  (c) and (d) In the nonstationary regime ($\Delta
  k=30\imm$, $\Iin=29~{\rm GW/cm^2}$), a 17
  fs pulse compressed pulse with trailing oscillations is observed. }
\label{fig:sr-nsr}
  \end{center}
\end{figure}

Experimentally, compression of mJ pulses was observed at near-infrared
wavelengths of 800 nm from 120 fs down to 30 fs when the compression
was done externally (using either a prism pair or a near-lossless block of
bulk calcite) \cite{liu:1999} and when working as a soliton compressor
from 120 fs to 45 fs \cite{ashihara:2002} and from 35 fs to 20 fs
\cite{moses:2005}. In an important recent advance, spatially uniform
compression using super-Gaussian flat-top beams was demonstrated also
at 800 nm \cite{moses:2007}. At 1260 nm compression to 12 fs (3
optical cycles) was achieved \cite{moses:2006}, while at telecom
wavelengths compression down to 35 fs was observed
\cite{ashihara:2004,zeng:2006}.  Numerical simulations have predicted
sub-2-cycle pulses, but it was clear already in the first experiments
that group-velocity mismatch (GVM) was a limiting factor for
compression\cite{liu:1999,moses:2006}.  It was observed that in the
so-called \textit{stationary regime} clean compression is possible,
while in the so-called \textit{nonstationary regime} GVM distorts the
compressed pulse too much to be of any practical use, and severe
reductions in compression capabilities is observed. As an example of
this, the numerical simulations in Figs.~\ref{fig:sr-nsr}(a)-(b) and
Figs.~\ref{fig:sr-nsr}(c)-(d) compare the pulse compression
performance under equal conditions in the two regimes. In the
stationary regime a 6 fs compressed pulse is observed while the
nonstationary regime the GVM effects are much stronger, resulting in a
17 fs compressed pulse with trailing oscillations.

Significant progress in understanding this was recently made by using
nonlocal theory (see Ref.\cite{krolikowski:2004} for a review on
optically nonlocal media). The GVM-induced Raman-like term found
previously \cite{ilday:2004,moses:2006} was shown to originate from a
temporally nonlocal response function \cite{bache:2007a}. The nonlocal
behaviour appears when approximating the phase-mismatched dispersive
SHG process in the cascading limit as a nonlinear convolution between
the FW and a nonlocal response function \cite{nikolov:2003}. An
accurate prediction of when the system is stationary or nonstationary
was presented, and the nonlocal theory predicted that an oscillatory
chirp was created on the FW for short enough pulses in the
nonstationary regime, which effectively limits the amount of
compression achievable and qualitatively explained the trailing
oscillations observed. It was also argued that the temporal time
scales of the nonlocal response function had an influence on the final
compressed pulse duration, but a systematic investigation was not
made.

On the other hand, a recent study showed that the performance of the
CQSC can conveniently be described by scaling laws involving an
effective soliton number $\neffs^2=\nshgb^2-\nkerrb^2$, appearing as
the difference between the quadratic and the cubic soliton numbers
\cite{bache:2007}. Since $\neffs$ depends only on input parameters
(such as parameters describing the input pulse intensity and duration
and the material), the compressed pulse properties
can be predicted using these scaling laws. Appropriate input
parameters can then be found giving compression to single-cycle
duration. However, neither the experiments nor the numerical
simulations have ever observed single-cycle compression, so obviously
there are higher order effects that prevent compression to reach such
levels. 
In particular, both experiments and simulations found optimal phase
mismatch values, where the best compression is observed. It is the
purpose of the present theoretical and numerical analysis to discover
what defines these optimal operation points.


\section{Propagation equations}
\label{sec:Prop-equat}

The SHG propagation equations in the slowly-evolving
wave-approximation (SEWA) are used to study pulses in a bulk quadratic
nonlinear crystal with single-cycle temporal resolution.  Neglecting
diffraction as well as the non-instantaneous cubic Raman
response the dimensionless equations for the FW ($\omega_1$) and
SH ($\omega_2=2\omega_1$) fields $U_{1,2}(\xi,\tau)$ are 
\cite{moses:2006b,bache:2007}
\begin{subequations}
\label{eq:shg-bulk}
\begin{align}
  \label{eq:shg-bulk-fh-no-kerr}
  &(i\partial_\xi 
  +\Dm_1') U_1
  +|\Delta \kzp|^{1/2}\nshgb \hat S_1'U_1^*U_2e^{i\Delta\kzp \xi}
+\nkerrb^2\hat S_1'U_1\left(|U_1|^2+B\bar{n}|U_2|^2\right)=0,\\
\nonumber
 &(i\partial_\xi-id_{12}'\partial_\tau+\Dm_{2,\rm eff}')
     U_2+|\Delta \kzp|^{1/2}\nshgb \hat S_2'U_1^2e^{-i\Delta\kzp \xi} 
  \\
  &\phantom{i\partial_\xi+}
+2\bar n^2\nkerrb^2 \hat S_2'U_2\left(|U_2|^2+B\bar n^{-1}|U_1|^2\right)=0.
  \label{eq:shg-bulk-sh-no-kerr}
\end{align}
\end{subequations}
Higher order dispersion (HOD) is included through the operator
$\Dm_j'\equiv \sum_{m=2}^{m_d} i^m
\delta_j^{(m)}\frac{\partial^m}{\partial \tau^m}$, with the
dimensionless dispersion coefficients $\delta_j^{(m)}\equiv
\kz_j^{(m)}(\tin^{m-2}|\bpp_1|m!)^{-1} $ and
$k_j^{(m)}\equiv\partial^m k_j/\partial \omega^m|_{\omega=\omega_j}$.
Since $k_j=n_j\omega_j/c$ is known analytically through the Sellmeier
equations of \cite{dmitriev:1999}, the exact dispersion
$\Dm_j=k_j(\omega)-(\omega-\omega_j)k_1^{(1)}-k_j(\omega_j)$ is used
in the numerics \cite{bache:2007}, corresponding to a dispersion order
$m_d=\infty$.  $n_j$ is the refractive index, and the phase mismatch
of the SHG process is $\Delta k=k_2-2k_1$. The Kerr cross-phase
modulation (XPM) term $B=2$ for type 0 SHG while for type I SHG
$B=2/3$ \cite{bache:2007}.  The time coordinate moves with the FW
group velocity $v_{\rm g,1}=1/\kp_1$, giving the GVM term
$d_{12}=v_{\rm g,1}^{-1}-v_{\rm g,2}^{-1}$. The equations are reported
in dimensionless form, $\tau=t/\tin$, where $\tin$ is the FW input
pulse duration, $\xi=z/\ld$, where $\ld=\tin^2/|\kpp_1|$ is the FW
dispersion length, and finally $U_1=E_1/\Ein$ and
$U_2=E_2/\sqrt{\bar{n}}\Ein$. Here $\Ein$ is the amplitude of the peak
electric input field, $\bar n=n_1/n_2$, $d_{12}'=d_{12}\tin/|\kpp_1|$,
and $\Delta \kzp= \Delta k\ld$. This scaling gives the quadratic (SHG)
and cubic (Kerr) soliton numbers \cite{moses:2006,bache:2007}
\begin{align}
  \nshgb^2=\frac{\ld\Ein^2\omega_1^2\deff^2}{c^2 n_{1}n_{2}|\Delta
  k|}, \quad \nkerrb^2=\frac{\ld n_{\rm
  Kerr,1}\Ein^2\omega_1}{c}
\end{align}
where $\deff$ is the effective quadratic nonlinearity, and $n_{{\rm
    Kerr},j}=3{\rm Re}(\chi^{(3)})/8 n_j$ is the cubic (Kerr)
nonlinear refractive index. $\nshgb$
might seem poorly defined in Eqs.~(\ref{eq:shg-bulk}) because of the
factor $|\Delta k'|^{1/2}$ in front of it, but the choice will become
clear later. Self-steepening is included through the operators $\hat
S_j'\equiv 1+i(\omega_j\tin)^{-1}\frac{\partial} {\partial \tau}$. The
combination of self-steepening and GVM implies that in the SEWA
framework the SH dispersion effectively is given by $\Dm_{2,\rm
  eff}'=\Dm_2'+\hat S_2'^{-1}\frac{\nu}{2}\frac{\partial^2
}{\partial\tau^2}$, where $ \nu\equiv cd_{12}^2/\omega_2n_2 |\bpp_1|$
\cite{moses:2006b,bache:2007}. We stress that all primed symbols in
our notation are the dimensionless form of the
corresponding unprimed symbol.

\section{Nonlocal model: reduced equation in the cascading limit}
\label{sec:Nonlocal-model}

\begin{figure}[tb]
  \begin{center}
    \centerline{
\includegraphics[width=6.2cm]{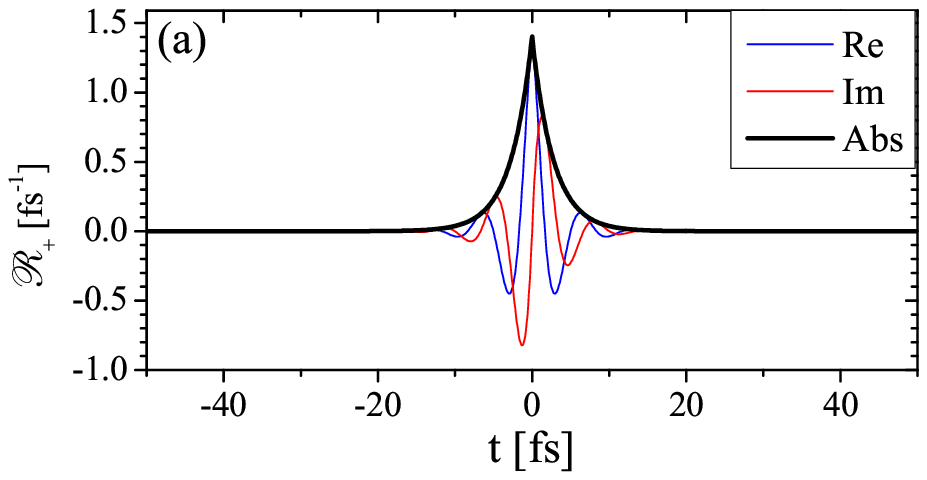}
\includegraphics[width=6.1cm]{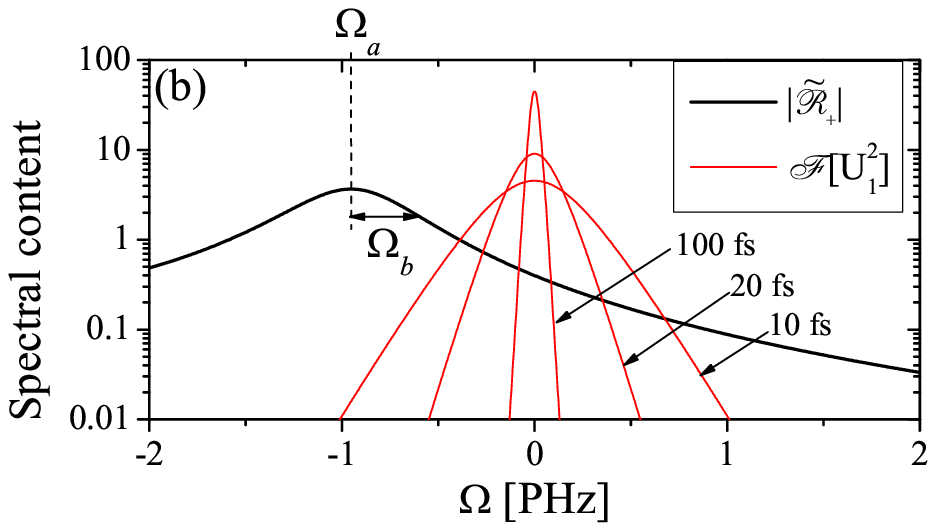}
}\centerline{
\includegraphics[width=6.2cm]{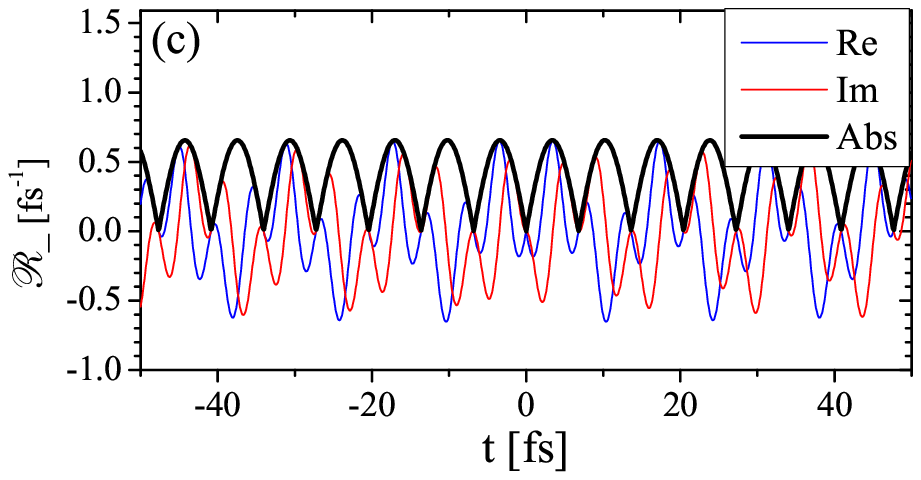}
\includegraphics[width=6.1cm]{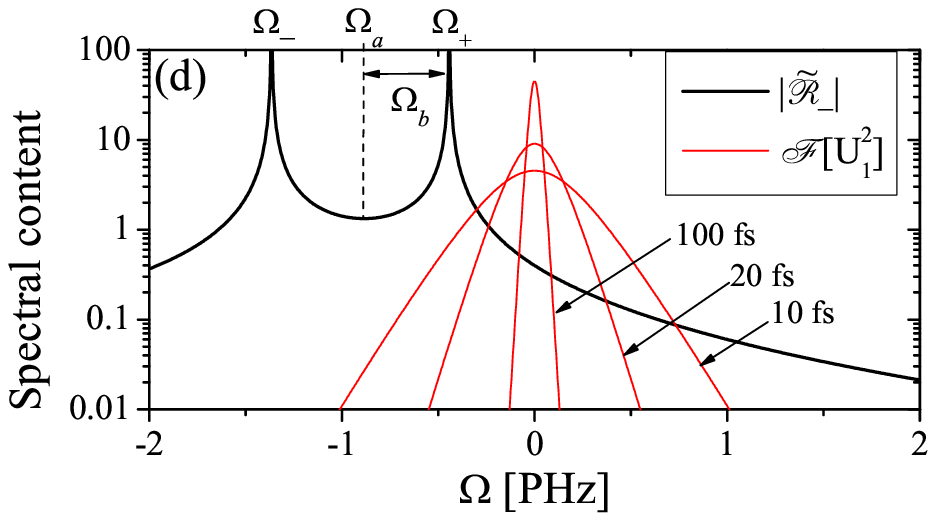}}
\caption{The nonlocal response functions in
  the (a,b) stationary regime ($s_b=1$) and (c,d) nonstationary regime
  ($s_b=-1$) as calculated for the simulation parameters in
  Fig.~\ref{fig:sr-nsr}. The cusp in the origin of $|\Rnl_\pm(t)|$ is
  typical for a Lorentzian response function. The spectral content of
  $U_1^2={\rm sech}^2(t/\tin)$ having 100, 20, and 10 fs FWHM duration
  is shown in (b,d). 
}\label{fig:Romega_nonstat}
  \end{center}
\end{figure}

The CQSC works in the cascading limit, \ie $|\Delta \kzp| \gg 1$, and
we henceforth take $\Delta \kzp>0$ as to have a self-defocusing
cascaded nonlinearity. In order to support solitons we then need
normal group-velocity dispersion (GVD), \ie $\kpp_j>0$. We now seek to
reduce the full SEWA model~(\ref{eq:shg-bulk}), in order to get some
physical insight into the compression process. In
Ref.~\cite{bache:2007a} it was shown that in the cascading limit
Eqs.~(\ref{eq:shg-bulk}) can be reduced to a single equation for the
FW
\begin{align}
  \label{eq:fh-shg-nlse-nonlocal}
  \left[i\frac{\partial}{\partial\xi}-
  \frac{1}{2}\frac{\partial^2}{\partial\tau^2}\right]U_1  
  +\nkerrb^2U_1|U_1|^2 
-\nshgb^2U_1^*\int_{-\infty}^{\infty} {\rm d}\tdummy
  R_\pm(\tdummy)U_1^2(\xi,\tau-\tdummy) =0.
\end{align}
In this derivation both self-steepening and \HOD were neglected
($\hat S_j=1$, $m_d=2$), but as shown later this can
straightforwardly be relaxed. Additionally the Kerr XPM
terms were neglected.  This dimensionless generalized nonlinear
Schr{\"o}dinger equation (NLSE) shows that the cascaded quadratic
nonlinearity imposes a temporal nonlocal response on the FW, governed
by the nonlocal response functions $R_\pm$ [examples are shown in
Fig.~\ref{fig:Romega_nonstat}(a,c)]. This model quantified the previous
qualitative definitions \cite{liu:1999,ilday:2004} of the stationary
and nonstationary regimes. When $d_{12}^2>2\Delta
kk_2^{(2)}$ the system is in the nonstationary regime and 
the oscillatory response function $R_-$ must be used. The criterion can be
interpreted using characteristic length scales, giving $
\lgvm^2<\lcoh L_{\rm D,2}/2\pi$; when GVM dominates its
length scale $L_{\rm GVM}=\tin/|d_{12}|$ becomes shorter than one
controlled by the product of the coherence length $\lcoh=\pi/|\Delta
k|$ and the SH GVD length scale $L_{\rm D,2}=\tin^2/|\kpp_2|$.  When $d_{12}^2<2\Delta kk_2^{(2)}$ the system is in the
stationary regime and one has to use the localized response function
$R_+$. We will now derive this result in details.

In the cascading limit $\Delta \kzp \gg 1$ the nonlocal approach
takes the ansatz
\begin{align}\label{eq:ansatz}
U_2(\xi,\tau)= \phi_2(\tau)\exp(-i\Delta\kzp\xi)  
\end{align}
This ansatz is assuming that all the dynamics in the propagation
direction of the SH is dominated by the phase mismatch, and the
condition for making this ansatz is that the coherence length
$\lcoh=\pi/\Delta \kz$ is much shorter than any other characteristic
length scale. This is true in the cascading limit except when the FW
is extremely short, in which case the GVM length $L_{\rm
  GVM}=\tin/|d_{12}|$ can become on the order of $\lcoh$.
Assuming $\nkerrb^2 U_2\ll \Delta \kzp^{1/2}\nshgb$ we may discard
the Kerr terms in Eq.~(\ref{eq:shg-bulk-sh-no-kerr}), and get an
ordinary differential equation (ODE)
\begin{align}\label{eq:ode}
  \delta_2^{(2)} \frac{d^2\phi_2}{d\tau^2}+id_{12}'
  \frac{d\phi_2}{d\tau}- \Delta\kzp \phi_2 =\Delta \kzp^{1/2}\nshgb
  U_1^2
\end{align}
where for simplicity we have only considered up to 2nd order
dispersion and neglected self-steepening and the SEWA correction to
the dispersion. We will come back to this point later. Fourier
transforming Eq.~(\ref{eq:ode}) and using the
Fourier transform pair $\tilde \phi_2(\Omega)=\FT[\phi_2](\Omega)\equiv
(2\pi)^{-1/2} \int_{-\infty}^\infty {\rm d}\tau e^{i\Omega \tau}
\phi_2(\tau)$ and $\phi_2(\tau)=\IFT[\tilde \phi_2](\tau)\equiv
(2\pi)^{-1/2}\int_{-\infty}^\infty {\rm d}\Omega e^{-i\Omega \tau}
\tilde \phi_2(\Omega) $ the ODE~(\ref{eq:ode}) becomes simply $
(\Omega^2\delta_2^{(2)}-\Omega d_{12}' + 
  \Delta\kzp)\tilde \phi_2(\Omega)=-\Delta \kzp^{1/2}\nshgb
  \FT[U_1^2](\Omega) $, 
so we can solve it in the frequency domain as
\begin{align}
\label{eq:Phi2}
\tilde \phi_2(\Omega)=-(2\pi/\Delta \kzp)^{1/2}\nshgb
\tilde R(\Omega)\FT[U_1^2](\Omega), \quad 
\tilde R(\Omega)\equiv \frac{\Delta \kzp(2\pi)^{-1/2}
  }{\delta_2^{(2)}\Omega^2-d_{12}'\Omega +
\Delta\kzp }
\end{align}
We now use the convolution theorem, 
so that in the time domain Eq.~(\ref{eq:Phi2}) becomes
\begin{align}
 \phi_2(\tau) = -\frac{\nshgb}{\sqrt{\Delta\kzp}}
 \int_{-\infty}^\infty {\rm d}\tdummy R(\tdummy) U_1^2(\xi,\tau-\tdummy)
\end{align}
where the nonlocal response function in the time domain is
$R(\tau)=\IFT[\tilde R(\Omega)]$. This result shows that in the
cascading limit the SH is affected by the FW in a nonlocal manner. 

The next step is to calculate $R(\tau)$. To do this it
is convenient to rewrite $\tilde R$  in Eq.~(\ref{eq:Phi2}) as
\begin{align}\label{eq:Qomega1}
  \tilde R(\Omega)=(2\pi)^{-1/2}\frac{\Omega_a'^2+s_b\Omega_b'^2}
  {(\Omega-\Omega_a')^2+s_b\Omega_b'^2} 
\end{align}
where we have introduced the dimensionless frequencies
and the sign parameters
\begin{subequations}
\begin{alignat}{2}
  \Omega_a'=&d_{12}'/2\delta_2^{(2)},
&\quad  \Omega_b'=&|\Delta \kzp/\delta_2^{(2)}-\Omega_a'^2|^{1/2}\\
  s_a=&{\rm sgn}[\Omega_a'],&\quad 
  s_b=&{\rm sgn}[\Delta \kzp/\delta_2^{(2)}-\Omega_a'^2]
\end{alignat}
\end{subequations}
The advantage of this notation is that it is now easy to characterize
the roots of the polynomial in the denominator, which is important
when we later will generalize to \HOD. 

When $s_b=+1$ Eq.~(\ref{eq:Qomega1}) becomes a Lorentzian centered in
$\Omega_a'$ and with the FWHM $2\Omega_b'$, see
Fig.~\ref{fig:Romega_nonstat}(b). The roots in the denominator of
Eq.~(\ref{eq:Qomega1}) are complex $\Omega=\Omega_a'\pm i\Omega_b'$.
The temporal response function, $R_+(\tau)$, can readily be calculated
by taking the inverse Fourier transform\footnote{Note: on
  dimensional form $\Rnl(t)=R(t/\tin)/\tin$, which is independent on
  $\tin$ since one must replace $\tau_{a,b}$ with the dimensional
  equivalent $t_{a,b}=\tau_{a,b} \tin$ in
  Eqs.~(\ref{eq:Rplus-dimless},\ref{eq:Rminus-dimless}). Instead in the
  frequency domain both $\tilde R$ and $\tilde \Rnl$ are
  dimensionless.}
\begin{align}\label{eq:Rplus-dimless}
R_+(\tau) &= \frac{{\ta}^2+{\tb}^2}{2\ta^2\tb} \exp(-is_a\tau/\ta)
\exp(-|\tau|/\tb) 
\end{align}
where we have introduced the dimensionless
characteristic nonlocal time scales
\begin{align}
  \ta=|\Omega_a'|^{-1}=|2\delta_2^{(2)}/d_{12}'|
,\quad\label{eq:tau1}
 \tb= \Omega_b'^{-1}=|\Delta \kzp/\delta_2^{(2)}-\Omega_a'^2|^{-1/2}
\end{align}
The localized nature of Eq.~(\ref{eq:Rplus-dimless}) is shown in
Fig.~\ref{fig:Romega_nonstat}(a), and $\tb$ controls the width of
$|R_+|$ while $\ta$ is the period of the phase oscillations. We note
that Eq.~(\ref{eq:Phi2}) is defined so 
$\int_{-\infty}^\infty {\rm d}\tau
R_+(\tau)=1$.

When $s_b=-1$, $\tilde R(\Omega)$ has two simple poles
at $\Omega=\Omega_a'\pm\Omega_b'\equiv \Omega_\pm'$, making  $\tilde
R(\Omega)$ diverge as shown in Fig.~\ref{fig:Romega_nonstat}(d). 
Then 
$R_-(\tau)=\IFT[\tilde R_-(\Omega)]$ exists as a
Cauchy principal value giving
\begin{align}\label{eq:Rminus-dimless}
R_-(\tau) &= \frac{{\ta}^2-{\tb}^2}{2\ta^2\tb} 
\exp(-is_a\tau/\ta) \sin(|\tau|/\tb)
\end{align}
In contrast to $R_+$, this response function 
is not localized but instead is oscillatory as a consequence of the
two poles in $\tilde R_-(\Omega)$ [see the example shown in
Fig.~\ref{fig:Romega_nonstat}(c)].

Now, using Eq.~(\ref{eq:Phi2}) with the ansatz~(\ref{eq:ansatz}) and
plugging it into Eq.~(\ref{eq:shg-bulk-fh-no-kerr}) we arrive at
Eq.~(\ref{eq:fh-shg-nlse-nonlocal}) under the aforementioned
approximations. This essentially shows the central result of
Ref.~\cite{bache:2007a}, namely that due to the cascaded quadratic
nonlinearity the FW can be described by a generalized NLSE with a
nonlocal temporal response in the Kerr-like SPM term. In Ref.
\cite{ilday:2004} the nonstationary regime was defined as when GVM
effects prevent the build-up of a nonlinear phase shift, which when
applied to soliton compression consequently results in poor
compression \cite{moses:2006}.  Based on the nature of the nonlocal
response functions, we can now clarify that the boundary to the
stationary regime is when $s_b$ changes sign, which in dimensional
units $\Delta k=\dksr$, where
\begin{align}\label{eq:stationary}
\dksr=\frac{d_{12}^2}{2k_2^{(2)}}
\end{align}
Thus, to be in the stationary regime the phase-mismatch must be
significantly large, $\Delta k>\dksr$. When GVM is weak compared to
the phase mismatch then $s_b=+1$, and the response function ($R_+$) is
monotonously decaying in magnitude: the convolution in the Kerr-like
SPM term in Eq.~(\ref{eq:fh-shg-nlse-nonlocal}) provides a finite
temporal response. Therefore this must correspond to the stationary
regime.  Instead when GVM is strong compared to the phase mismatch
then $s_b=-1$, and the response function ($R_-$) is oscillating and
non-decaying: the temporal response from the convolution is no longer
finite. Thus, this must correspond to the nonstationary regime.  

Finally, to include the influence of self-steepening in the nonlocal
theory, Eq.~(\ref{eq:fh-shg-nlse-weakly-nonlocal}) would have the
operator $\hat S_1'$ acting on all nonlinear terms and
Eq.~(\ref{eq:ode}) would have the operator $\hat S_2'$ acting on the
RHS. In frequency
domain 
we may thus use a steepening-corrected response function $\tilde
R^{\rm ss}(\Omega)\equiv \tilde R (\Omega)[1+(\omega_2
\tin)^{-1}\Omega]$.  This does not change the transition value $\Delta
k_{\rm sr}$. Lastly, the NLS-like nonlocal
equation~(\ref{eq:fh-shg-nlse-weakly-nonlocal}) will have the operator
$\hat S_1'$ acting on all nonlinear terms. It should also be stressed that
self-steepening can affect the Raman-like term in
Eq.~(\ref{eq:fh-shg-nlse-weakly-nonlocal}) \cite{moses:2006b}, but
this effect does not appear in the nonlocal model used here because
this would require taking into account higher-order perturbation terms
[\ie making a more elaborate ansatz than Eq.~(\ref{eq:ansatz})].

\section{The weakly nonlocal limit}
\label{sec:weakly-nonl-limit}

The nonlocal response in Eq.~(\ref{eq:fh-shg-nlse-nonlocal}) can be
better understood in the weakly nonlocal limit, where the width of the
nonlocal response function is much narrower than the width of $U_1^2$.
The resulting simplified equation gives a better physical insight
\cite{krolikowski:2000}, and is important because it governs the
initial dynamics (until pulse compression makes $U_1$ so short that
the nonlocal response is no longer weak). We evaluate the convolution
in the frequency domain $\int_{-\infty}^\infty {\rm d}\tdummy
R(\tdummy)
U_1^2(\xi,\tau-\tdummy)  
=\int_{-\infty}^\infty {\rm d}\Omega e^{-i\Omega \tau}\tilde
R(\Omega)\FT[U_1^2](\Omega)$ for convenience.
In the weakly nonlocal limit $\tilde R(\Omega)$ is approximated by a
1st order expansion around $\Omega=0$, where $\FT[U_1^2](\Omega)$ is
non-vanishing.  This holds when $\tilde R(\Omega)$ varies slowly
compared to $\FT[U_1^2](\Omega)$.  In this case 
\begin{align}\label{eq:Rtilde-omega-expand}
\tilde R(\Omega)\FT[U_1^2](\Omega)\simeq \left[\tilde R(\Omega=0)+\Omega
  \frac{{\rm d}\tilde
R}{{\rm d} \Omega}|_{\Omega=0}\right]\FT[U_1^2]  (\Omega)
\end{align}
which in the time domain equivalently is
$R(s)U_1^2(\xi,\tau-\tdummy)\simeq
R(s)\left[U_1^2(\xi,\tau)-\tdummy\partial U_1^2(\xi,\tau)/\partial
  \tau\right]$. However, in the nonstationary regime the integration
is done over two simple poles located on the $\Omega$-axis. Using the
residue theorem the frequency integral $\int_{-\infty}^\infty {\rm
  d}\Omega e^{-i\Omega \tau}\tilde R(\Omega)\FT[U_1^2](\Omega)$ can be
evaluated as a contour integral, giving a contribution from the Cauchy
principal value of the integral, and a contribution from deforming the
integration contour around the poles on the real $\Omega$-axis.  The
residual contributions from the poles to the frequency integral are
\cite{bache:2008a}
\begin{align}\label{eq:residuals}
\rho(\tau,U_1)&=
i{\rm
  sgn}(\tau)\sqrt{\frac{\pi}{2}}
  \frac{\ta^2-\tb^2}{2\ta^2\tb}
\left[e^{-i\tau\Omega_+'}\FT[U_1^2](\Omega_+')-
  e^{-i\tau\Omega_-'}\FT[U_1^2](\Omega_-') 
  \right]  
\end{align}
which consist of an oscillatory component in form of complex
exponentials with frequencies $\Omega'_\pm$ each weighted by the
spectral strength of $U_1^2$ at that frequency. Thus, the
influence of this contribution becomes important when the FW is short
enough for its spectrum to cover the range, where $\Omega_\pm'$ are
located, see Fig.~\ref{fig:Romega_nonstat}(d). Using
Eq.~(\ref{eq:residuals}) and $(2\pi)^{1/2}\tilde R(0)=1$ and
$(2\pi)^{1/2}{\rm d} \tilde R/{\rm d}
\Omega|_{\Omega=0}=2\Omega_a'/(\Omega_a'^2+s_b \Omega_b'^2)$, the
nonlocal convolution is
\begin{align}
\int_{-\infty}^\infty {\rm
  d}\tdummy R(\tdummy) 
U_1^2(\xi,\tau-\tdummy)
&\simeq
U_1^2+is_a\tau_{R,\rm SHG}U_1\frac{\partial
  U_1}{\partial \tau}+\frac{1-s_b}{2}\rho(\tau,U_1)
\label{eq:Rtilde-expand}
\end{align}
Now introducing the effective soliton number
$\neffs^2=\nshgb^2-\nkerrb^2$, Eq.~(\ref{eq:fh-shg-nlse-nonlocal})
becomes
\begin{multline}
  \label{eq:fh-shg-nlse-weakly-nonlocal}
  \left[i\frac{\partial}{\partial\xi}-
  \frac{1}{2}\frac{\partial^2}{\partial\tau^2}\right]U_1  
  -\neffs^2 U_1|U_1|^2 
=\nshgb^2\left[is_a\tau_{R,\rm SHG}|U_1|^2\frac{\partial
  U_1}{\partial \tau}+\frac{1-s_b}{2}U_1^*\rho(\tau,U_1)\right]
\end{multline}
The first term on the RHS is a GVM-induced Raman-like perturbation
caused by the cascaded SHG nonlinearity. It is Raman-like due to the
asymmetry of $R_\pm$ \cite{bache:2007a}, stemming from the phase term
$\exp(-is_a\tau/\ta)$ in Eqs.~(\ref{eq:Rplus-dimless})
and~(\ref{eq:Rminus-dimless}).  It has the characteristic
dimensionless time $\tau_{R, \rm SHG}\equiv
4|\Omega_a'|/(\Omega_a'^2+s_b\Omega_b'^2)=2|d_{12}'|/\Delta k'$
\cite{ilday:2004,moses:2006,bache:2007a,bache:2007}, which on dimensional
form reads
\begin{align}\label{eq:trSHG}
  T_{R, \rm SHG}=\tau_{R, \rm SHG}\tin=2|d_{12}|/\Delta k
\end{align}
The direct dependence on the GVM-parameter $d_{12}$ implies that the
Raman-like perturbation disappears in absence of GVM.
Equation~(\ref{eq:fh-shg-nlse-weakly-nonlocal}) was also derived for
the stationary regime $s_b=1$ in Ref.\cite{bache:2007a}\footnote{The
  factor $s_a$ on the RHS of
  Eq.~(\ref{eq:fh-shg-nlse-weakly-nonlocal}) was unfortunately lost
  during the proofs in Eq. (12) of Ref. \cite{bache:2007a}.},
obviously without the contribution $U_1^*\rho(\tau,U_1)$.

Eq.~(\ref{eq:fh-shg-nlse-weakly-nonlocal}) is a very strong result: It
states that in the weakly nonlocal limit the effective soliton number
$\neffs$ can be used in the scaling laws of \cite{bache:2007} to
predict, e.g., the optimal compression point.  Previously these
scaling laws were thought to hold only in the stationary regime
\cite{bache:2007a,bache:2007}.  The result also tells us that in the
weakly nonlocal limit, a central observation of Ref. \cite{moses:2006}
is confirmed: for a given, fixed value of $\Delta k$, the Raman-like
effect of the first term of the RHS of
Eq.~(\ref{eq:fh-shg-nlse-weakly-nonlocal}) becomes increasingly
significant with increasing $\nshgb^2$, thus limiting the possible
compression ratio.  However, it is now clear that in the nonstationary
regime, the Raman-like distortion is accompanied by an oscillatory
perturbation term $U_1^*\rho(t,U_1)$ which also increases with
$\nshgb^2$.  In both the stationary and nonstationary regimes the
Raman-like distortions place a limitation on the maximum soliton
order, but in the nonstationary regime both terms on the RHS of
Eq.~(\ref{eq:fh-shg-nlse-weakly-nonlocal}) distort the compression,
and the combined effect is more severe, a result that is also
confirmed by numerical simulations below.  On the other hand, it is
also obvious that for low soliton numbers these detrimental effects
are weak and it is actually possible to observe clean compressed
pulses even in the nonstationary regime, as observed by both numerical
simulations as well as experiments
\cite{ashihara:2002,moses:2005}.\footnote{These experiments were
  actually done in the nonstationary regime according to the nonlocal
  theory.}. This is also apparent in Eq.~(\ref{eq:residuals}) where
the oscillations can be neglected if $\FT[U_1^2](\Omega_\pm')$ is
vanishing, \ie if the compressed pulse duration is long enough.

In the stationary regime the characteristic time scale that decides
the strength of the nonlocal response is $\tbp=\tb\tin$. The weakly
nonlocal approximation applies when $\tbp\ll \Delta t$, where $\Delta
t$ is the FW pulse duration. But when does it apply in the
nonstationary regime? We know that the width $\Delta \Omega$ of
$\FT[U_1^2]$ is $\Delta \Omega\propto \Delta t^{-1}$.  Referring to
Fig.~\ref{fig:Romega_nonstat}(d) we must require that the positions of
the two poles $\Omega_\pm$ be sufficiently far away from the frequency
range, where $\FT[U_1^2]$ is nonvanishing, \ie $|\Omega_\pm|\gg \Delta
\Omega$. In physical units this implies that the weakly nonlocal limit
in the nonstationary regime can be expressed by the requirement
$\Delta t \gg \tap \tbp/|\tap-\tbp|$.

Let us evaluate this requirement. It is important to notice from
Eq.~(\ref{eq:tau1}) that $\tbp$ diverges at the transition $\Delta
k_{\rm sr}$, see also Fig.~\ref{fig:Neff=8}(a). Thus, in the
nonstationary regime the requirement $\Delta t \gg \tap
\tbp/|\tap-\tbp|$ holds even for quite short pulses as long as $\tap$
and $\tbp$ are not too similar. This is generally true close to the
transition $\Delta k_{\rm sr}$, while away from the transition
$\tbp\simeq \tap$ because $\Delta k$ gets small, see
Eq.~(\ref{eq:tau1}) and Fig.~\ref{fig:Neff=8}(a). In this case we can
no longer be sure to be in the weakly nonlocal limit. In the
stationary regime the system will initially be in the weakly nonlocal
limit $\Delta t \gg \tbp$ except close to the transition $\Delta
k_{\rm sr}$, where $\tbp$ diverges. It is interesting to notice that
the weakly nonlocal limit in the stationary regime is well away from
the transition $\Delta k_{\rm sr}$, while in the nonstationary regime
it is close to the transition $\Delta k_{\rm sr}$. We finally remark
that $\tap$ also may diverge when GVM is negligible. This implies that
the factor $e^{-i s_a\tau/\ta}= 1$, so $R(\tau)$ becomes real and
symmetric. Thus, the 1st order correction on the RHS of
Eq.~(\ref{eq:fh-shg-nlse-weakly-nonlocal}) disappears because the
Raman-like perturbation vanishes ($T_{R, \rm SHG}=0$) and a 2nd order
correction must be made. 

\section{Numerical results and discussion}
\label{sec:Numer-results-disc}

We will now apply the results of the nonlocal theory to understand
realistic numerical simulations of Eq.~(\ref{eq:shg-bulk}). It is
important to stress that the theory neglects Kerr XPM effects and that the
coherence length is the shortest length scale in the system. This
latter requirement implies that the system can \textit{initially} be
well described by the nonlocal theory, but as the pulse is compressed
the GVM (and other length scales) can become so short that this is no
longer true. Therefore the accuracy of the nonlocal model will not
always be enough to accurately predict the outcome of the numerical
simulations and the experiments. However, since the nonlocal model
often will be an adequate approximation for a large part of the
propagation through the nonlinear medium, we can still use it to
understand the temporal dynamics until that potentially happens.


We will now show that there are two main categories of compression
limitations.

\begin{enumerate}
\item Effects limiting phase-mismatch range where
  compression is possible and efficient
  \begin{enumerate}
  \item In the nonstationary regime $\Delta k<\Delta k_{\rm sr}$
    the oscillatory nonlocal response function implies that 
    compression is inefficient unless the soliton order is very low. 
  \item Competing cubic nonlinearities pose an upper limit $\Delta k_c$
    \cite{bache:2007a,bache:2007} beyond which $\neffs<1$ always.
    Close to this limit detrimental cubic XPM effects are observed.
  \end{enumerate}
\item Effects limiting the compression for a selected phase-mismatch
  value
  \begin{enumerate}
  \item The effective soliton order $\neffs
    =(\nshgb^2-\nkerrb^2)^{1/2}$ controls in the
    weakly nonlocal limit the compression factor
    $f_C=\tin/\topt=4.7(\neffs-0.86)$ \cite{bache:2007}.
  \item Nonlocal effects. 
    In the stationary regime
    $\topt$ is limited by the strength of the nonlocal response function
    $\tbp$. In the nonstationary regime $\topt$ is limited by the
    characteristic time $T_{R,\rm SHG}$ of the GVM-induced Raman-like
    perturbation.
  \item Propagation effects pertaining solely to the FW, such as
    higher-order dispersion, the Raman effect (negligible in nonlinear
    crystals) and cubic self-steepening.
  \item Competing cubic nonlinearities necessitate large quadratic soliton
    orders $\nshgb$, which increases detrimental nonlocal effects such
    as the Raman-like perturbation.
\end{enumerate}
\end{enumerate}


The numerical simulations of Eqs.~(\ref{eq:shg-bulk}) were performed
using a $\beta$-barium-borate crystal (BBO) as the quadratic nonlinear
medium.  The phase mismatch was changed through angle-tuning of the
crystal in a type I SHG configuration (implying $B=2/3$, see
\cite{bache:2007} for further details), and we are interested in
$\Delta k>0$, for which GVD is normal and $d_{12}<0$ (so
$s_a<0$).

\begin{figure}[tb]
  \begin{center}
    \centerline{\includegraphics[height=6.cm]{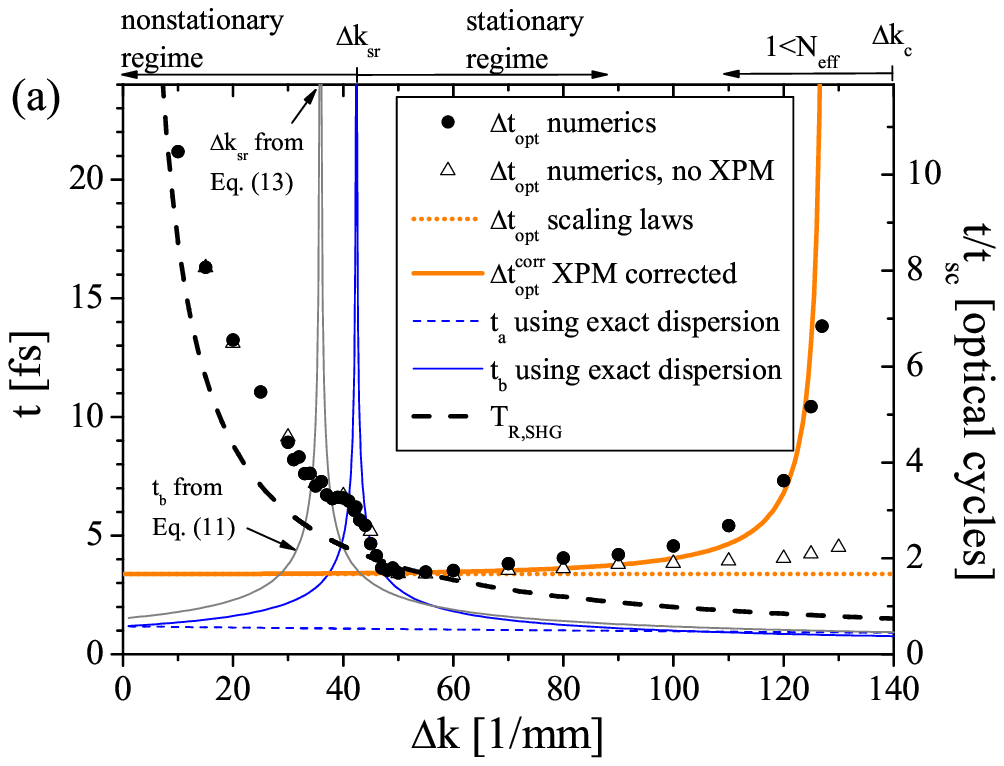}
    \includegraphics[height=5.5cm]{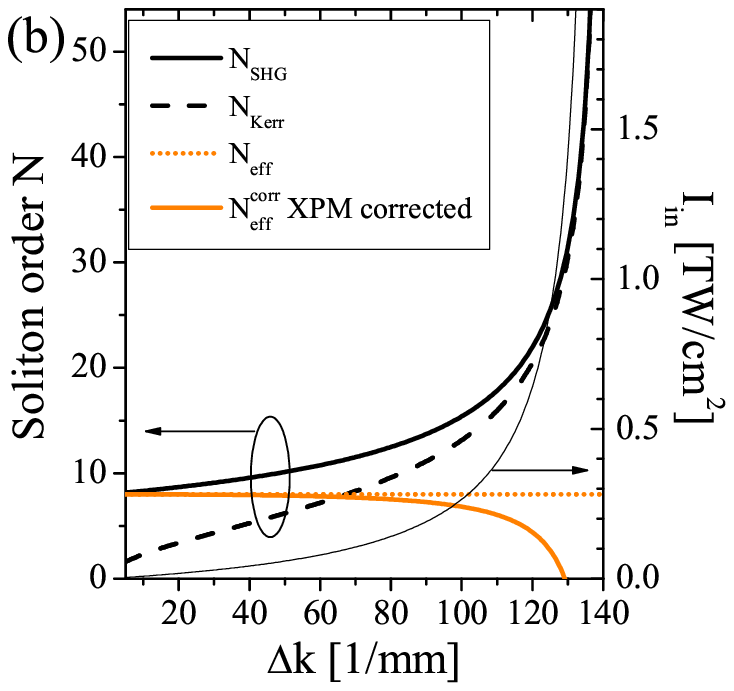}}
\caption{Data from numerical simulations as in
  Fig.~\ref{fig:sr-nsr} for different values of $\Delta k$. (a) The FW
  duration $\topt=\topt^{\rm FWHM}/1.76$ at $z=z_{\rm opt}$ is
  shown both for the full SEWA model~(\ref{eq:shg-bulk}), and when
  neglecting the Kerr XPM terms.  The lines show the
  nonlocal time scales $t_{a,b}=\tin\tau_{a,b}$, the characteristic
  Raman-like time $T_{R,\rm SHG}=2|d_{12}|/\Delta k$, and the
  predicted $\topt$ from the scaling laws \cite{bache:2007} as well as
  the predicted $\toptc$ when correcting for XPM effects on $\neffs$.
  $t_b$ as calculated using only up to second-order dispersion ($m_d=2$,
  gray curve) is also shown. The right ordinate shows time
  normalized to the single-cycle pulse duration $t_{\rm
    sc}=2.0$~fs.  (b) The SHG and Kerr soliton numbers required to
  have $\neffs=8$ fixed, achieved by adjusting $\Iin$.  The
  corrected effective soliton number due to XPM effects $\neffs^{\rm
    corr}$ is also shown. }\label{fig:Neff=8}
  \end{center}
\end{figure}

\begin{figure}[t]
  \begin{center}
    \centerline{\includegraphics[width=6.50cm]{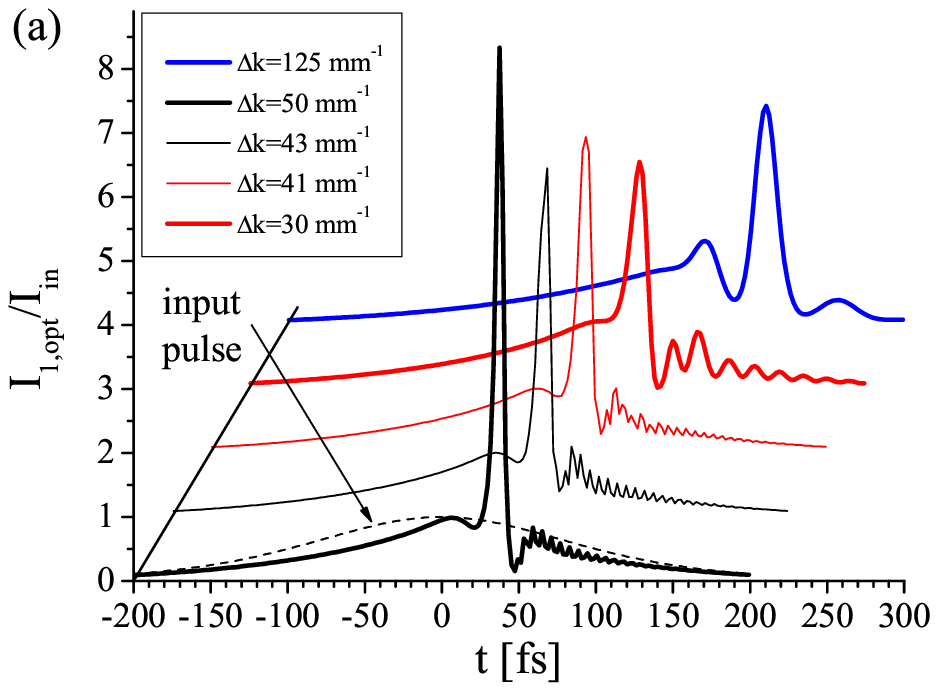}
\includegraphics[width=6.4cm]{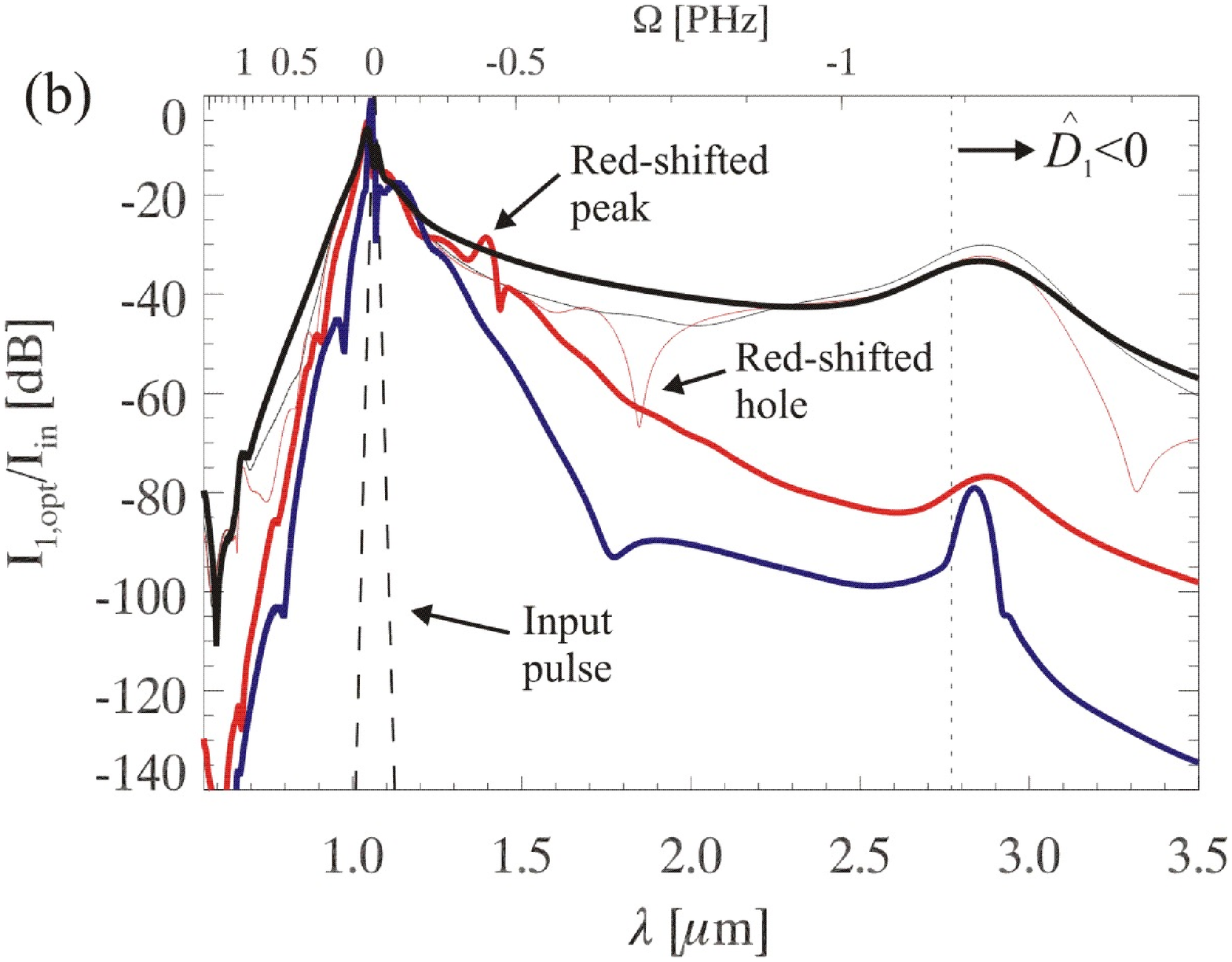}
}  \centerline{
\includegraphics[width=6.40cm]{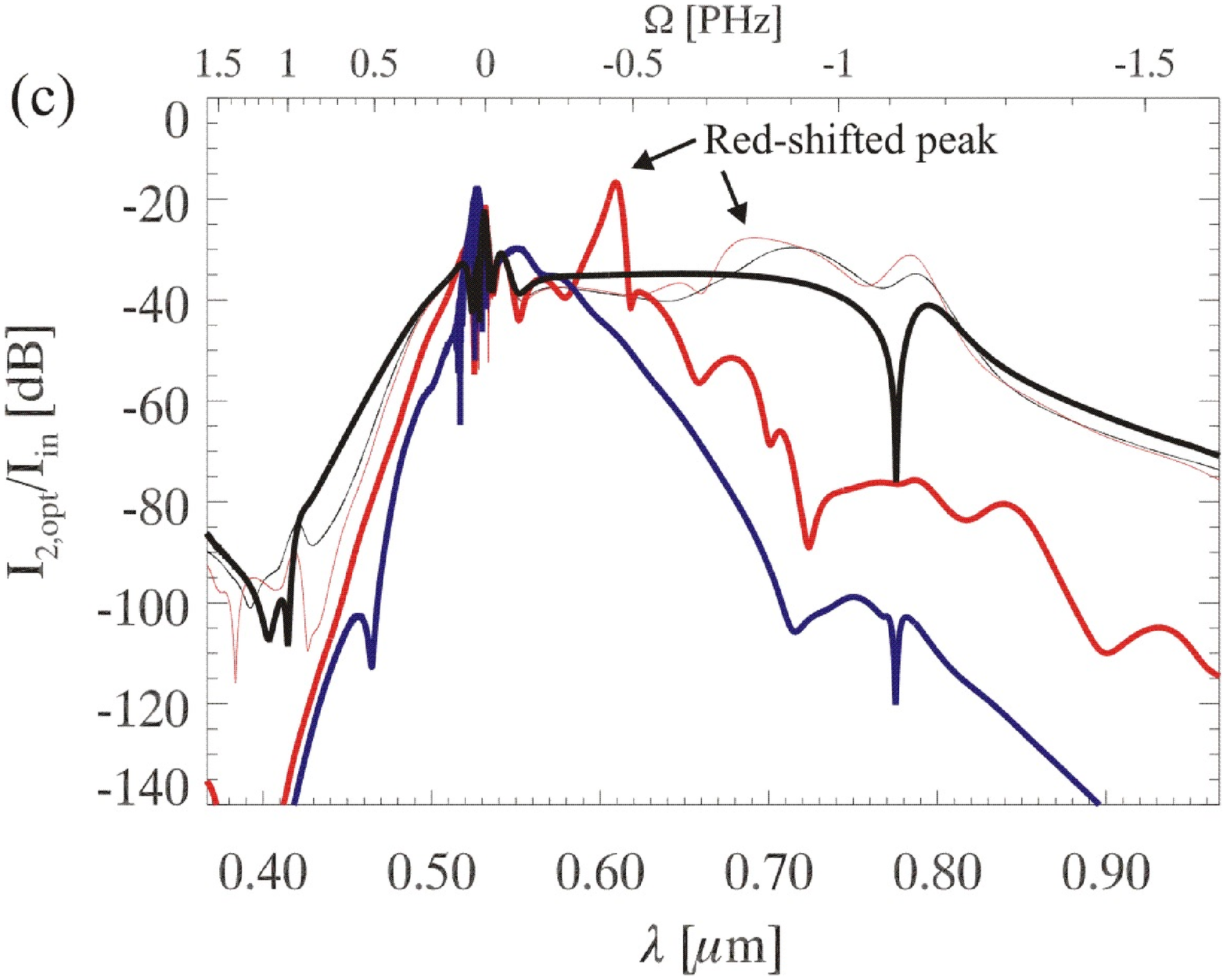}
\includegraphics[width=6.5cm]{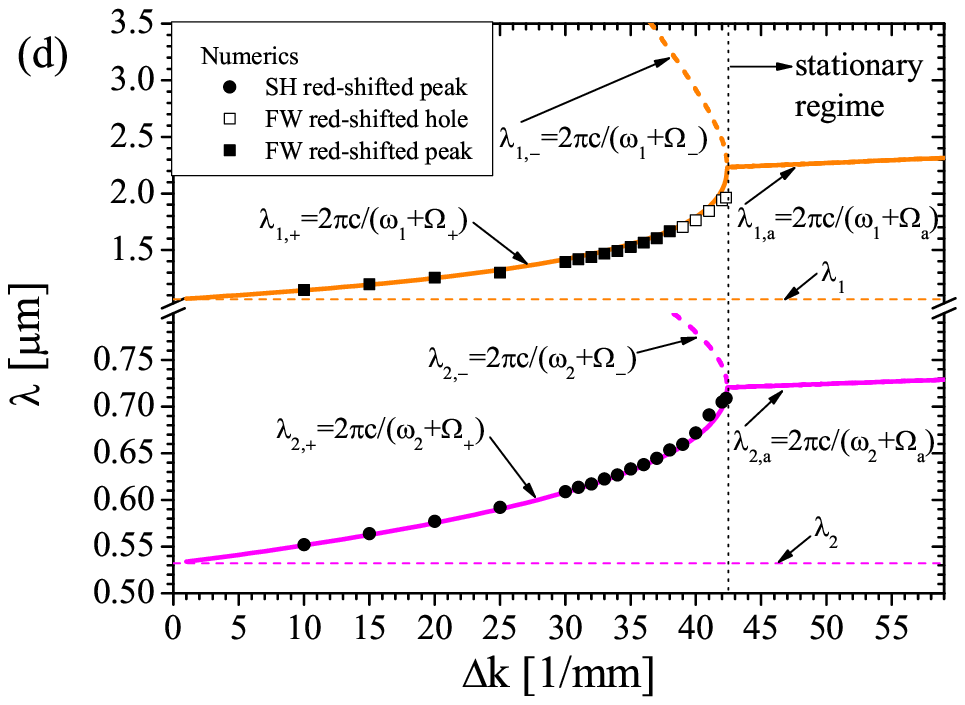}}
\caption{Data from the numerical results in Fig.~\ref{fig:Neff=8} for
  selected values of $\Delta k$: (a) $|U_1|^2$ at $z=z_{\rm opt}$
  versus time, (b) the corresponding FW and (c) SH wavelength spectra.
  Only up to $\lambda=3.5~{\rm \mu m}$ is shown in (b) since this is
  the edge of the transparency window of BBO \cite{dmitriev:1999}. The
  plots for $\Delta k=50\imm$ and $30\imm$ are the cuts in
  Fig.~\ref{fig:sr-nsr} at $z=z_{\rm opt}$.  (d) The position of the
  red-shifted spectral peaks in the nonstationary regime. The symbols
  are the results of numerical calculations while the lines are the
  predictions of the nonlocal theory. }\label{fig:Neff=8-spectra}
  \end{center}
\end{figure}

Figure~\ref{fig:Neff=8} summarizes simulations of pulse compression of
a 200 fs FWHM pulse centered at $\lambda_1=1064$ nm: the FW pulse
duration at the point of optimal compression $\Delta t_{\rm opt}$
(dark circles) is plotted as the phase mismatch $\Delta k$ is sweeped.
The strength of the cascaded quadratic nonlinearity
$\nshgb^2\propto\Delta k^{-1}$, while the Kerr nonlinearity remains
unchanged. However, in the plot we keep $\neffs=8$ fixed by adjusting
the input intensity $\Iin=\tfrac{1}{2}n_1\varepsilon_0
c\Ein^2$.\footnote{This is a typical experimental situation, because
  the optimal compression point $z_{\rm opt}$ scales with $\neffs$
  \cite{bache:2007}, and since the nonlinear crystal length is a
  constant parameter one adjusts the intensity so $z_{\rm opt}$
  coincides with the crystal length.} This implies that the scaling
law \cite{bache:2007} predicts equal compression everywhere (dotted
orange line).  While we do observe such a compression in the regime
around $\Delta k=50\imm$ [this example is shown in
Fig.~\ref{fig:sr-nsr}(a)-(b), and also in
Fig.~\ref{fig:Neff=8-spectra}], away from this point the compression
becomes sub-optimal.

Before explaining the results in detail, we must mention that the
nonlocal time scales plotted in Fig.~\ref{fig:Neff=8}(a) are not those
of Eq.~(\ref{eq:tau1}); these only take into account up to
second-order dispersion (dispersion order $m_d=2$), while in the numerical
simulations the dispersion is calculated exactly from the Sellmeier
equations \cite{bache:2007}, and subsequently corrected in the SEWA
framework up to 30th order \cite{bache:2007}. This poses a correction
on the nonlocal time scales as well as the transition to the
nonstationary regime, which was calculated numerically: we replaced
the polynomial in the denominator of $\tilde R$ in Eq.~(\ref{eq:Phi2})
with $\Dm_{2,{\rm eff}}$ (evaluated in frequency domain). The
transition to the nonstationary regime happens when a root-pair
switches from being each others complex conjugate to being purely real
and nondegenerate. Then $\Omega_{a,b}$ and $t_{a,b}$ can be extracted
from these roots. The transition to the stationary
regime~(\ref{eq:stationary}) is now simply found when $t_b$ diverges.
Fig.~\ref{fig:Neff=8} shows for comparison also $t_b$ calculated with
$m_d=2$, \ie using Eq.~(\ref{eq:tau1}).

The degrading compressor performance observed for large $\Delta k$ in
Fig.~\ref{fig:Neff=8}(a) is caused by the onset of XPM-effects. As
$\Delta k$ is increased the cascaded quadratic nonlinearity is
reduced, so in order to keep $\neffs=8$ the input intensity $\Iin$
must be increased, see Fig.~\ref{fig:Neff=8}(b).  Eventually the
required intensity diverges because $\Delta k$ approaches the
so-called upper limit of the compression window $\Delta k_c$, beyond
which $\neffs<1$ always \cite{bache:2007a,bache:2007}. As $\Iin$
becomes large so does the input beam fluence $\Phi_{\rm
  in}=2\tin\Iin$, and this makes XPM effects more pronounced; as shown
in \cite{bache:2007} above a critical fluence of $\Phi_c=33~{\rm
  mJ/cm^2}$ the onset of compression in a BBO is not $\neffc=1$ -- as
one would expect from the NLS-like
equation~(\ref{eq:fh-shg-nlse-nonlocal}) -- but can approximately be
described by the scaling law $\neffc=1+\Delta \neffs$, with $\Delta
\neffs=\Phi_{\rm in}/[\Phi_c(1+\Phi_c/\Phi_{\rm in})]$ as the delay in
onset. The delay is caused by the XPM term creating an
intensity dependent self-focusing phase-shift in addition to the one
already created by the Kerr SPM term. The immediate consequence is
that we can no longer expect the compression factor of $f_C=33.5$
predicted from $\neffs=8$; instead we must use a corrected effective
soliton number $\neffs^{\rm corr}=\neffs-\Delta \neffs$ [see
Fig.~\ref{fig:Neff=8}(b)], which for high fluences then will give a
reduced compression performance.  Figure~\ref{fig:Neff=8}(a) shows
$\topt$ as predicted from the scaling laws for $\neffs=8$ (dotted
orange line), together with the corrected $\toptc$ as calculated using
$\neffs^{\rm corr}$ (solid orange line).  As expected for high $\Delta
k$ values $\toptc$ starts to deviate from $\topt$.  Importantly, it
seems to describe very accurately the compression performance observed
numerically. An example of how the pulse looks like in this regime is
shown in Fig.~\ref{fig:Neff=8-spectra} ($\Delta k=125\imm$, blue
curve). The compressed pulse is longer (18 fs FWHM) and we also
checked that it compresses later than what one would expect with
$\neffs=8$. These values correspond very well to what the reduced
soliton number $\neffs^{\rm corr}\simeq 3$ predicts through the
scaling laws. Thus, XPM strongly degrades compression when the beam
fluence becomes large.  This is also confirmed by the simulations
shown with open circles in Fig.~\ref{fig:Neff=8}(a), for which XPM
effects were turned off: only a weak degradation in compression is
seen for high $\Delta k$.

For $\Delta k=43-50\imm$ the limit to compression is determined by the
strength of the nonlocal response function. Close to the transition to
the nonstationary regime $t_b$ becomes large, so the nonlocal response
$\Rnl_+$ is very broad. Initially, however, the 200 fs FWHM input
pulse sees only a weakly nonlocal response. As the pulse compresses
the nonlocal response becomes strongly nonlocal, whereby the NLS-like
model~(\ref{eq:fh-shg-nlse-nonlocal}) reduces to a linear
Schr{\"o}dinger equation having a potential defined by the response
function
\cite{snyder:1997,krolikowski:2000,PhysRevE.64.016612,Shadrivov:2002,nikolov:2003}.
The pulse cannot be narrower than the width of this
potential given by $t_b$, which explains the behaviour observed for
phase-mismatch values just above $\dksr$. This has also been
observed for spatial nonlocal solitons
\cite{snyder:1997,krolikowski:2000,PhysRevE.64.016612,Shadrivov:2002,nikolov:2003}.

For $\Delta k<42\imm$ the system is in the nonstationary regime and as
$\Delta k$ is reduced $\Delta t_{\rm opt}$ increases as $\Delta
k^{-1}$. The compression limit does therefore not follow $\tbp$, but
instead follows the characteristic Raman-like time $T_{R,\rm
  SHG}=2|d_{12}|/\Delta k$ quite closely.  Indeed some physical
explanation can be extracted from this parameter, since it namely
represents the pulse duration, where the GVM length $\lgvm= \Delta
t/|d_{12}|$ becomes shorter than the coherence length
$\lcoh=\pi/\Delta k$. Intuitively it seems logical that the compressed
pulse duration hit a limit when the GVM length is equal to the
coherence length: the cascaded nonlinear interaction can no longer
build up the phase shift because the GVM will remove the FW and SH
from each other before even one cascaded cycle is complete. It is
interesting to note that arguments similar to these were initially
used to define the nonstationary regime \cite{liu:1999,ilday:2004},
and it was already there clear that compression was limited in the
same way as shown here. These studies were, as mentioned before, carried
out in the nonstationary regime as defined by the nonlocal analysis,
so the results corroborate each other.  Another important result from
the nonlocal analysis is that there is an oscillatory contribution in
the nonstationary regime $\rho(\tau,U_1)$, see
Eq.~(\ref{eq:residuals}). When we consider a pulse that gets shorter
and shorter, the strength of $\FT[U_1^2](\Omega_\pm)$ becomes larger
and larger [see Fig.~\ref{fig:Romega_nonstat}(d)], and so we start in the
present case to observe oscillations with frequency
$\Omega_+=\Omega'_+/\tin$ (since $\Omega_+$ in the present example is
the frequency closest to origin).  That means that the soliton sees an
oscillating potential with period $\propto |\Omega_+|^{-1}$, so we
would expect that the pulse duration can never go below $
|\Omega_+|^{-1}$.  This explains very well why the pulse duration
increases in the nonstationary regime when $\Delta k$ is decreased. It
should also be mentioned that in the nonstationary regime $T_{R,\rm
  SHG}=|\Omega_+^{-1}+\Omega_-^{-1}|$, which therefore supports this
explanation. Note that no data points below $\Delta k=10\imm$ are
shown because the cascading limit breaks down there
\cite{desalvo:1992}.
  
To better understand the difference between the stationary and
nonstationary regimes, Fig.~\ref{fig:Neff=8-spectra}(a) shows examples
of compressed pulses.  For $\Delta k=50\imm$ a 6.3 fs FWHM compressed
pulse is observed as expected from the scaling laws, while as the
transition to the nonstationary regime is approached ($\Delta
k=43\imm$) the pulse compression is limited by the nonlocal potential
of strength $\tbp$.  Once inside the nonstationary regime, the pulse
not only becomes compressed poorly, but trailing oscillations are
evident. The corresponding FW and SH wavelength-spectra are shown in
Fig.~\ref{fig:Neff=8-spectra}(b) and (c). For $\Delta k=50\imm$ both
the FW and SH spectra are very flat, except for a spectral FW peak and
corresponding spectral SH hole. As we explain below these peaks are
actually dispersive waves. Closer to the transition ($\Delta
k=43\imm$) the SH spectrum develops a peak because $\tilde
\Rnl_+(\Omega)$ here is a very narrow Lorentzian. Inside the
nonstationary regime a distinct red-shifted peak grows up in the SH
spectrum, which can be explained by the nonlocal theory since the
spectral peak sits at the frequency $\Omega_+$. In turn, close to the
transition ($\Delta k=41\imm$) the FW has a corresponding spectral
hole at $\omega_1+\Omega_+$, while further from the transition
($\Delta k=30\imm$) it becomes a spectral peak. To confirm this, we
show in Fig.~\ref{fig:Neff=8-spectra}(d) the red-shifted holes/peaks
found numerically versus $\Delta k$, with an impressive agreement with
the nonlocal theory. This FW spectral hole/peak is the main limitation
to the pulse compression in the nonstationary regime.


The FW spectra in Fig.~\ref{fig:Neff=8-spectra}(b) have pronounced
peaks around $\lambda=2.9~\mu{\rm m}$. To understand this we observed
that the FW dispersion operator in the frequency domain
$\Dm_1(\Omega)=\sum_{m=2}^\infty m!^{-1}\Omega^m\kz_1^{(m)}$ changes
sign and for $\lambda_1=1.064~\mu{\rm m}$ becomes negative beyond the
dotted line in Fig.~\ref{fig:Neff=8-spectra}. Based on previous
results from the NLSE
\cite{PhysRevA.51.2602,skryabin:2003,cristiani:2003} at this point the
compressed soliton should therefore be phase matched to a linear
dispersive wave. Such dispersive waves have not been observed before
with cascaded quadratic nonlinearities, but their appearance further
underlines the analogy between soliton compression with cascaded
quadratic nonlinearities and with cubic nonlinear media.  The
wavelength of the dispersive wave $\lambda_{\rm dw}$ seems not to
change as $\Delta k$ is varied, which is due to the FW dispersion
being independent on the crystal angle. Instead $\lambda_{\rm dw}$
changes strongly with the soliton frequency, but the soliton-frequency
blue shifts observed in the simulations were quite small and similar
(around 30-40 THz).  The solitons are blue shifted because $\Delta
k>0$ and $s_a<0$ and the shifts explain why the dispersive waves are
slightly red-shifted compared to the dotted line. The dispersive waves
can also be noted in Fig.~\ref{fig:sr-nsr}(b,d). They do not emerge
before the pulse is compressed, because their strength is related to
the spectral strength of the soliton to which they are coupled
\cite{cristiani:2003}. 

\begin{figure}[t]
  \begin{center}
\centerline{
\includegraphics[height=3.8cm]{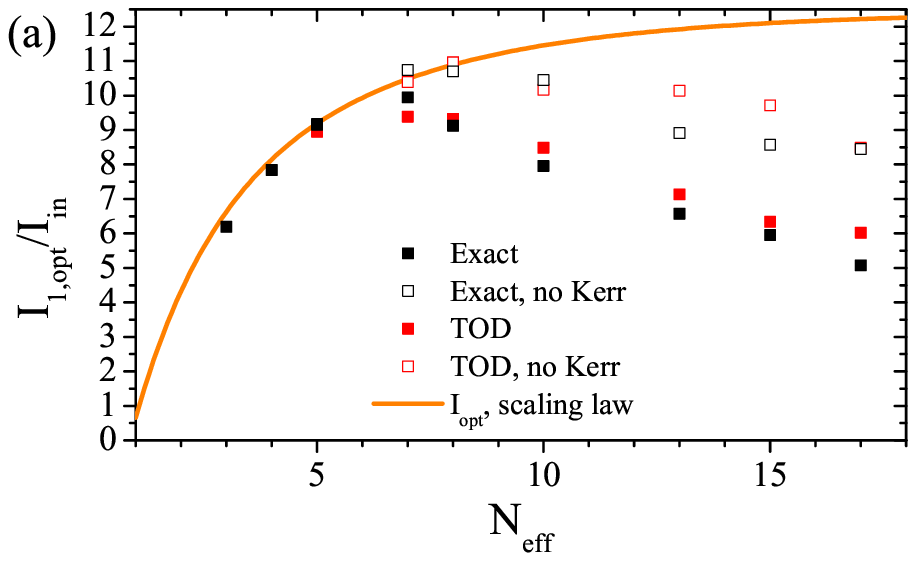}
\includegraphics[height=3.8cm]{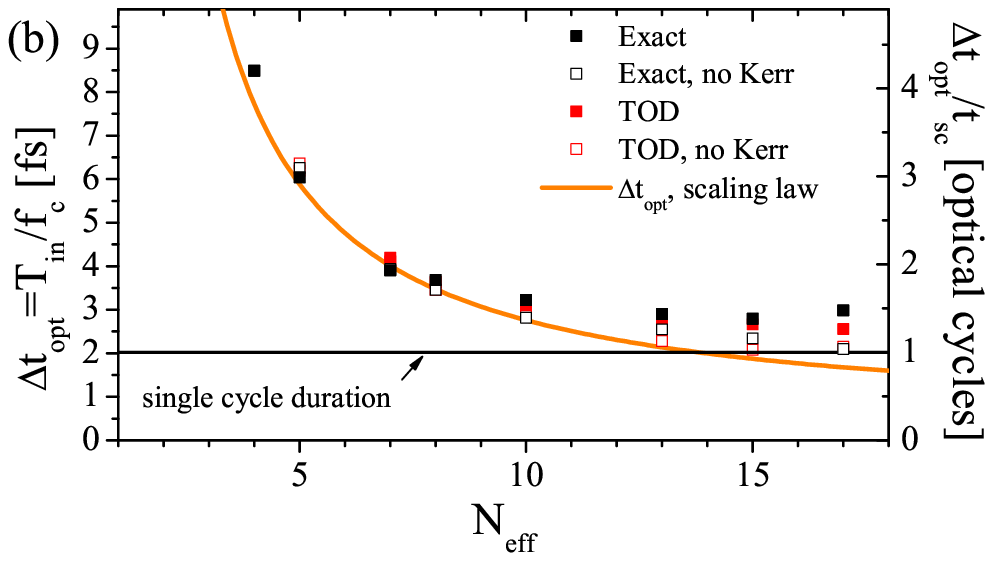}}
\centerline{
\includegraphics[width=6.5cm]{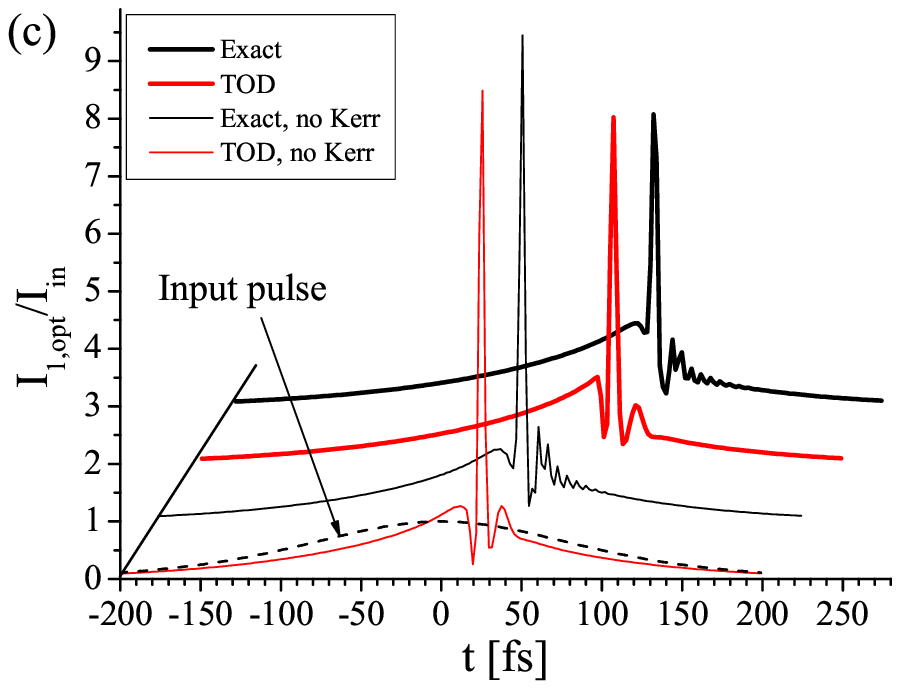}
\includegraphics[width=6.5cm]{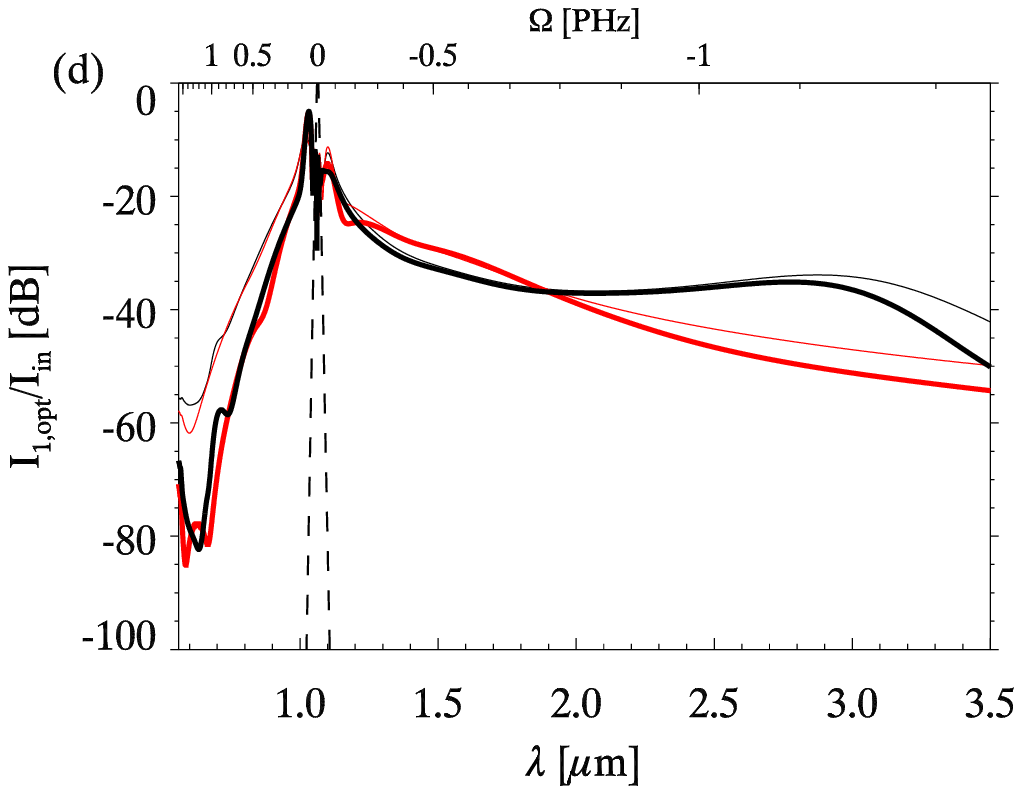}}
\caption{Results of pulse compression simulations as in Fig.~\ref{fig:Neff=8},
  taking $\Delta k=60\imm$ and varying $\neffs$. (a) and (b) show
  $I_{1,\rm opt}/\Iin$ and $\Delta t_{\rm opt}$ when using exact
  dispersion ($m_d=\infty$) and when including up to TOD ($m_d=3$), as
  well as the same simulations without competing Kerr nonlinearities.
  The orange curves are the predicted values from the scaling laws
  \cite{bache:2007}, corrected for XPM. (c) and (d) show FW time and
  spectral profiles for $\neffs=17$.  }\label{fig:1064-neff}
  \end{center}
\end{figure}

What happens when the effective soliton order in the stationary regime
is pushed to create single-cycle pulses? In a previous study it was
found that the GVM-induced Raman-like perturbation beyond some optimal
soliton order starts to dominate and makes the compressed pulse
asymmetric, while the peak intensity drops (Fig. 1 in Ref.
\cite{moses:2006}, where $\Delta k=16~\pi/{\rm mm}$). These results
are confirmed in Fig.~\ref{fig:1064-neff}(a), showing the peak
intensity of the compressed pulse versus $\neffs$ for $\Delta
k=60\imm$. Beyond $\neffs=8$ the pulse compression deviates
from the prediction of the scaling law, even for the
simulations including only up to third order dispersion (TOD, \ie
dispersion order $m_d=3$), or neglecting the competing Kerr
nonlinearities. However, the compressed pulse duration $\topt$, shown
in Fig.~\ref{fig:1064-neff}(b), decreases even beyond this point of
maximum intensity. The explanation is that $\Delta t_{\rm
  opt}=\tin/f_c$ in (b) is determined by the compression factor $f_c$
alone, while the intensity in (a) is $I_{\rm 1,opt}/\Iin=f_cQ_c$,
where $Q_c$ is the compressed pulse quality relating the energy of the
central spike to the initial pulse energy. Around $\neffs=17$ the
compressed pulse is quite close to the single-cycle regime. The time
profiles for this case are shown In Fig.~\ref{fig:1064-neff}(c). The
simulations without Kerr nonlinearities (so $\nshgb=\neffs=17$)
actually predict single-cycle compressed pulses, while turning on the
Kerr nonlinearities the compressed pulses are no longer single-cycle
but increase to 1.5 optical cycles. Notice also that the pulses with
Kerr nonlinearities are more asymmetric. This asymmetry is caused by
the GVM-induced Raman-like perturbations [1st term on the RHS of
Eq.~(\ref{eq:fh-shg-nlse-weakly-nonlocal})] as pointed out in
Ref.\cite{moses:2006}. Since this effect stems from the quadratic
nonlinearities it must be stressed that they are also affecting the
single-cycle pulses obtained without Kerr nonlinearities. However, the
difference is that there $\nshgb=17$ while with Kerr nonlinearities
$\nshgb=22.9$ must be chosen to have $\neffs=17$. Therefore the
strength of the Raman-like perturbation, which scales as $\nshgb^2$,
is much stronger when including the competing Kerr nonlinearities
leading to a more asymmetric pulse. We also note that the simulations
with exact dispersion (no polynomial expansion, $m_d=\infty$) have
fast trailing oscillations, which are absent for the TOD simulations.
The FW spectra in Fig.~\ref{fig:1064-neff}(d) offer an explanation:
only with exact dispersion is there a spectral peak around 3 $\mu$m;
these are phase-matched dispersive waves.  With TOD the spectrum is
instead smooth because the phase-matching condition for the dispersive
wave is pushed far into the infra-red.

Considering these results, a brief discussion on the effect on TOD in
soliton compressors is fruitful. In fiber soliton compressors the most
detrimental effect on pulse compression is the Raman self-scattering
term, that comes from a non-instantaneous Kerr nonlinear response. As
mentioned before, this effect can be neglected in nonlinear crystals.
However, Chan and Liu showed that TOD can severely distort the pulse
when GVD is small \cite{chan:1995}, while for larger GVD the TOD is
less important for the pulse shape \cite{chan:1998}, and simply leads
to a slowing down of the soliton. In this context the TOD effect
observed here, namely that the phase-matching wavelength of the
dispersive wave is shifted, is completely different. We also remark
that in our simulations TOD is positive, and GVD is not small
($\kpp_1\simeq 40~{\rm fs^2/mm}$ and $\kpp_2\simeq 100~{\rm fs^2/mm}$).

Summing up, the GVM-induced Raman-like distortions prevents
efficient compression at high soliton numbers (as previously found in
Ref.~\cite{moses:2006}), and the competing cubic
nonlinearities aggravates this effect. In addition, the presence of a
dispersive wave prevented soliton compression to the single-cycle
level. The dispersive wave disappears when only including up to TOD
because the dispersion is no longer accurate in the region where the
dispersive wave is observed, and this underlines the importance of
including \HOD in the numerics. An interesting case to study would be
when a large dispersion control is possible, such as with photonic
crystal fibers \cite{bache:2005a}. One could namely imagine that the
dispersion could be engineered, so the phase-matching condition for
the dispersive wave is pushed into the far IR-regime, allowing for
further compression towards the single-cycle regime.

\section{Conclusions}
\label{sec:Conclusions}

To summarize we have shown that the limits to compression in cascaded
quadratic soliton compressors can in most cases accurately be
understood from a nonlocal model, which describes the cascaded
quadratic nonlinearity as a nonlocal Kerr-like self-phase modulation
response. 

In the stationary regime, where the nonlocal response is localized,
one cannot compress pulses beyond the width (strength) of the nonlocal
response function. When increasing the effective soliton order to
potentially compress beyond single-cycle duration, the numerical
simulations indicated that competing Kerr nonlinear effects were
preventing single-cycle compressed pulses: Since the quadratic soliton
number must be chosen much larger than without Kerr nonlinearities,
this increases detrimental effects such as the GVM-induced Raman-like
perturbation found using the nonlocal theory. Additionally it was
found that higher-order dispersion can also prevent the observation of
single-cycle compressed pulses. In particular, dispersive waves
phase-matched to the compressed higher-order soliton caused trailing
oscillations on the compressed pulse, eventually impeding further
compression even at higher intensities.

In the nonstationary regime the nonlocal response function is
oscillatory, and the convolution between the pulse and the nonlocal
response function in the weakly nonlocal limit can elegantly be
understood in the same way as in the stationary regime, except for an
oscillatory contribution that causes trailing oscillations in the
compressed pulse and severely degrades compression. The SH spectrum
was found to be strongly red-shifted to a wavelength accurately
predicted by the nonlocal theory. This spectral shift in turn induces
a peak in the FW, which is the main compression limitation in the
nonstationary regime. The compression limit was found to be the
characteristic Raman-like response time of the cascaded process
$T_{R,\rm SHG}$, roughly the pulse duration for which the GVM and the
coherence lengths become identical.

Another compression limit is set by the material Kerr nonlinearity, 
which makes compression possible only below a critical phase mismatch
parameter, and requires large soliton orders for successful
compression. Thus, higher-order effects (XPM, higher-order dispersion
and self-steepening) come into play and detrimental nonlocal effects
are increased.  The XPM effects were accurately accounted for by using
a corrected (reduced) soliton number.

The present analysis will serve as a useful tool for further
experimental progress in soliton compression using cascaded quadratic
nonlinearities. We will now focus our attention to compression in a
BBO at $\lambda_1=800$~nm, because GVM is much stronger than what was
presented here. Thus we expect the nonlocal analysis to provide more
insight into this case, in particular concerning the nonstationary
regime, which is the dominating one at 800 nm.

M.B. acknowledges support from The Danish Natural Science Research
Council (FNU, grant no.  21-04-0506).

\end{document}